\documentclass[sort&compress,final]{aipproc}\layoutstyle{6x9}
\begin{document}
\title{Heavy-meson decay constants from QCD sum rules}
\classification{11.55.Hx, 12.38.Lg, 03.65.Ge}
\keywords{Nonperturbative QCD, sum rules, beauty and charmed
mesons}
\author{Wolfgang Lucha}{address={HEPHY, Austrian Academy of
Sciences, Nikolsdorfergasse 18, A-1050 Vienna, Austria}}\iftrue
\author{Dmitri Melikhov$^{*,}$}{address={Institute of Nuclear
Physics, Moscow State University, 119991, Moscow, Russia }}
\author{Silvano Simula}{address={INFN, Sezione di Roma 3, Via
della Vasca Navale 84, I-00146, Roma, Italy}}\fi

\begin{abstract}
We sketch a recent sum-rule extraction of the decay constants of
the heavy pseudoscalar mesons $D$, $D_s$, $B$, and $B_s$ from the
two-point correlator of heavy--light pseudoscalar currents
\cite{lms2010}.~Our main emphasis lies on the control over all the
uncertainties in the decay constants, related both~to~the input
QCD parameters and to the limited accuracy of the method of sum
rules. Gaining this control has become possible by application of
our new procedure of extracting hadron observables based~on a dual
threshold depending on the Borel parameter. For the charmed-meson
decay constants,~we~find \begin{eqnarray*}f_D&=&(206.2\pm
7.3_{(\rm OPE)}\pm 5.1_{(\rm syst)})\;\mbox{MeV},\\
f_{D_s}&=&(245.3\pm 15.7_{(\rm OPE)}\pm 4.5_{(\rm
syst)})\;\mbox{MeV}.\end{eqnarray*} For the beauty mesons, the
decay constants turn out to be extremely sensitive to the precise
value~of the $\overline{\rm MS}$ mass of the $b$-quark,
$\overline{m}_b(\overline{m}_b)$. By requiring our sum-rule
estimate to match the average of~the lattice determinations
of~$f_B$, we extract the rather accurate value
$$\overline{m}_b(\overline{m}_b)=(4.245\pm 0.025)\;\mbox{GeV}.$$
Feeding this parameter value into our sum-rule formalism leads to
the beauty-meson decay~constants
\begin{eqnarray*}f_{B} &=& (193.4 \pm 12.3_{\rm (OPE)} \pm 4.3_{\rm
(syst)})\; {\rm MeV},\\ f_{B_s} &=& (232.5 \pm 18.6_{\rm (OPE)}
\pm 2.4_{\rm (syst)})\; {\rm MeV}.\end{eqnarray*}\vspace*{-2ex}
\end{abstract}

\pacs{11.55.Hx, 12.38.Lg, 03.65.Ge}
\maketitle

\section{Introduction}
The extraction of ground-state decay constants from
Shifman--Vainshtein--Zakharov sum rules \cite{svz} is a
complicated problem: First, one should construct a reliable
operator product expansion (OPE) for the Borel-transformed
correlation function $\Pi(\tau)$, where $\tau$ denotes the Borel
parameter, of two pseudoscalar heavy--light currents. We make use
of the OPE for this correlator to three-loop accuracy
\cite{chetyrkin}, reshuffled in terms of the $\overline{\rm MS}$
heavy-quark masses, in which case the perturbative expansion
exhibits a reasonable convergence \cite{jamin}.

Second, even if the parameters of this OPE are known precisely,
the knowledge of the truncated OPE for a correlator allows one to
extract the characteristics of the bound state only with some
error which reflects the intrinsic uncertainty of the method of
QCD sum rules. Acquiring control over this systematic uncertainty
poses a very subtle problem~\cite{lms_2ptsr}.

Recently, we have formulated a novel approach for extracting
ground-state parameters from some correlator \cite{lms_new}. Let
us briefly recall the most essential features of our approach: As
consequence of the assumption of quark--hadron duality the
ground-state contribution and the OPE with a cut applied at some
effective continuum threshold $s_{\rm eff}$ are related by
\begin{eqnarray}
\label{SR_QCD} f_Q^2 M_Q^4 e^{-M_Q^2\tau}=\Pi_{\rm dual}(\tau,
s_{\rm eff}(\tau)) \equiv \int\limits^{s_{\rm
eff}(\tau)}_{(m_Q+m)^2} ds \, e^{-s\tau}\rho_{\rm pert}(s) +
\Pi_{\rm power}(\tau).
\end{eqnarray}
Here, $\rho_{\rm pert}(s)$ is the perturbative spectral density
known to order $\alpha_s^2$; $\Pi_{\rm power}$ describes the
series of power corrections expressed in terms of condensates of
increasing dimensions.

Of course, in order to extract the decay constant one has to fix
the effective continuum threshold $s_{\rm eff}$. Moreover, as is
obvious from (\ref{SR_QCD}), $s_{\rm eff}$ should be a function of
$\tau$. Otherwise the $\tau$-dependences of the l.h.s.\ and the
r.h.s.\ of (\ref{SR_QCD}) do not match each other. However, the
{\em exact effective threshold\/}, corresponding to employing on
the l.h.s.\ of (\ref{SR_QCD}) the exact~hadron mass and decay
constant, is clearly not known. The extraction of hadron
parameters from the sum rule consists therefore in attempting (i)
to find a good approximation to the exact continuum threshold and
(ii) to acquire control over the accuracy of this approximation.

Let us introduce the dual invariant mass $M_{\rm dual}$ and the
dual decay constant $f_{\rm dual}$ by the definitions
\begin{eqnarray}
\label{mdual} M_{\rm dual}^2(\tau)&\equiv&-\frac{d}{d\tau}\log
\Pi_{\rm dual}(\tau, s_{\rm eff}(\tau)),
\\
\label{fdual} f_{\rm dual}^2(\tau)&\equiv&M_Q^{-4}
e^{M_Q^2\tau}\Pi_{\rm dual}(\tau, s_{\rm eff}(\tau)).
\end{eqnarray}
If the mass of the ground state is known, any deviation of this
dual mass from the actual ground-state mass yields an indication
of the excited-state contributions picked up by~the dual
correlator. Assuming a specific functional form of the effective
continuum threshold and requiring the least deviation of the dual
mass (\ref{fdual}) from the known ground-state mass in the Borel
window leads to a variational solution for the effective
threshold. As soon~as this latter quantity has been fixed, one
calculates the decay constant from (\ref{fdual}). The (na\"ive)
standard assumption for the effective threshold is that it is just
a $\tau$-independent constant. In addition to this approximation,
we have considered polynomials in $\tau$: the reproduction of the
dual mass improves considerably for the $\tau$-dependent
quantities, which means~that a dual correlator with
$\tau$-dependent threshold isolates the ground-state
contribution~much better and is less contaminated by excited
states than a dual correlator with the traditional
$\tau$-independent threshold. As consequence, the accuracy of
extracted hadron observables improves considerably. Experience
gained from the study of potential models shows~that the band of
values obtained from the linear, quadratic, and cubic Ans\"atze
for the effective threshold encompasses the true value of the
decay constant \cite{lms_new}. It was also shown that~the
extraction procedures in quantum theory and QCD are quantitatively
very similar~to~each other \cite{lms_qcdvsqm}. This talk
summarizes our recent findings \cite{lms2010} on heavy-meson decay
constants.

\section{Decay constants of the $D$ and $D_s$ mesons}
\begin{figure}[!ht]
\begin{tabular}{cc}
\includegraphics[height=6.5cm]{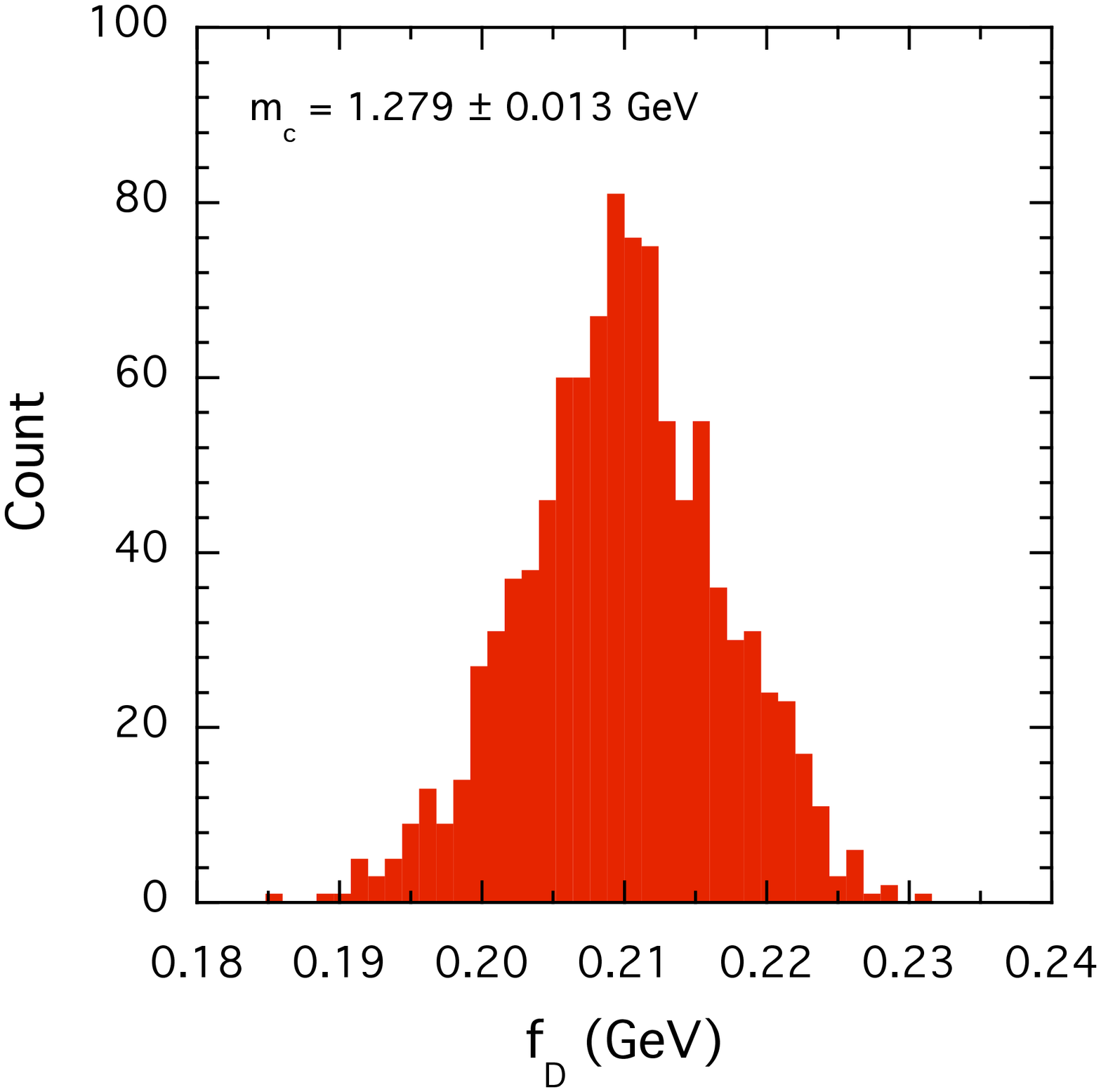}&
\includegraphics[height=6.5cm]{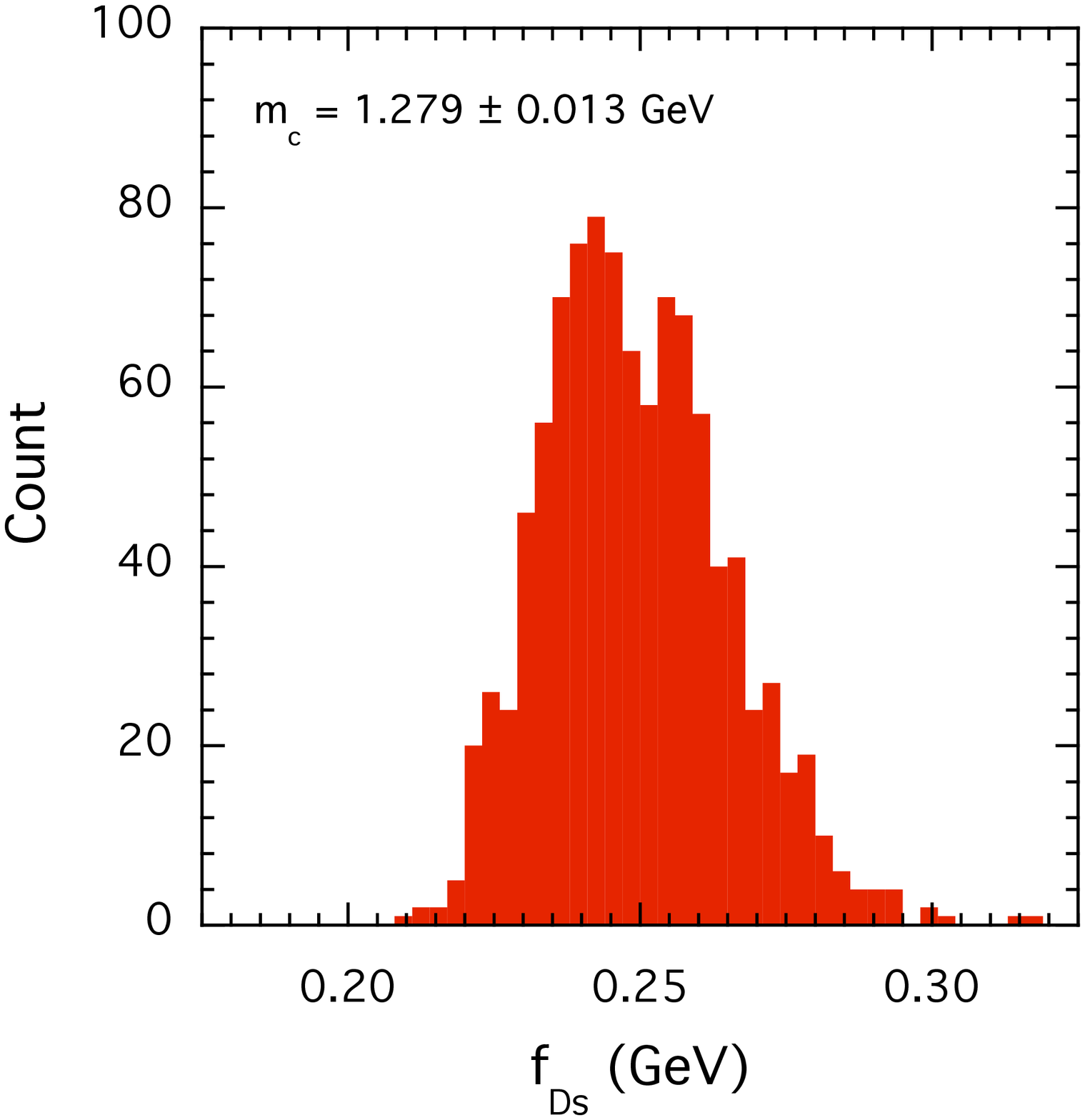}
\\
(a) & (b)
\\
\includegraphics[height=6.5cm]{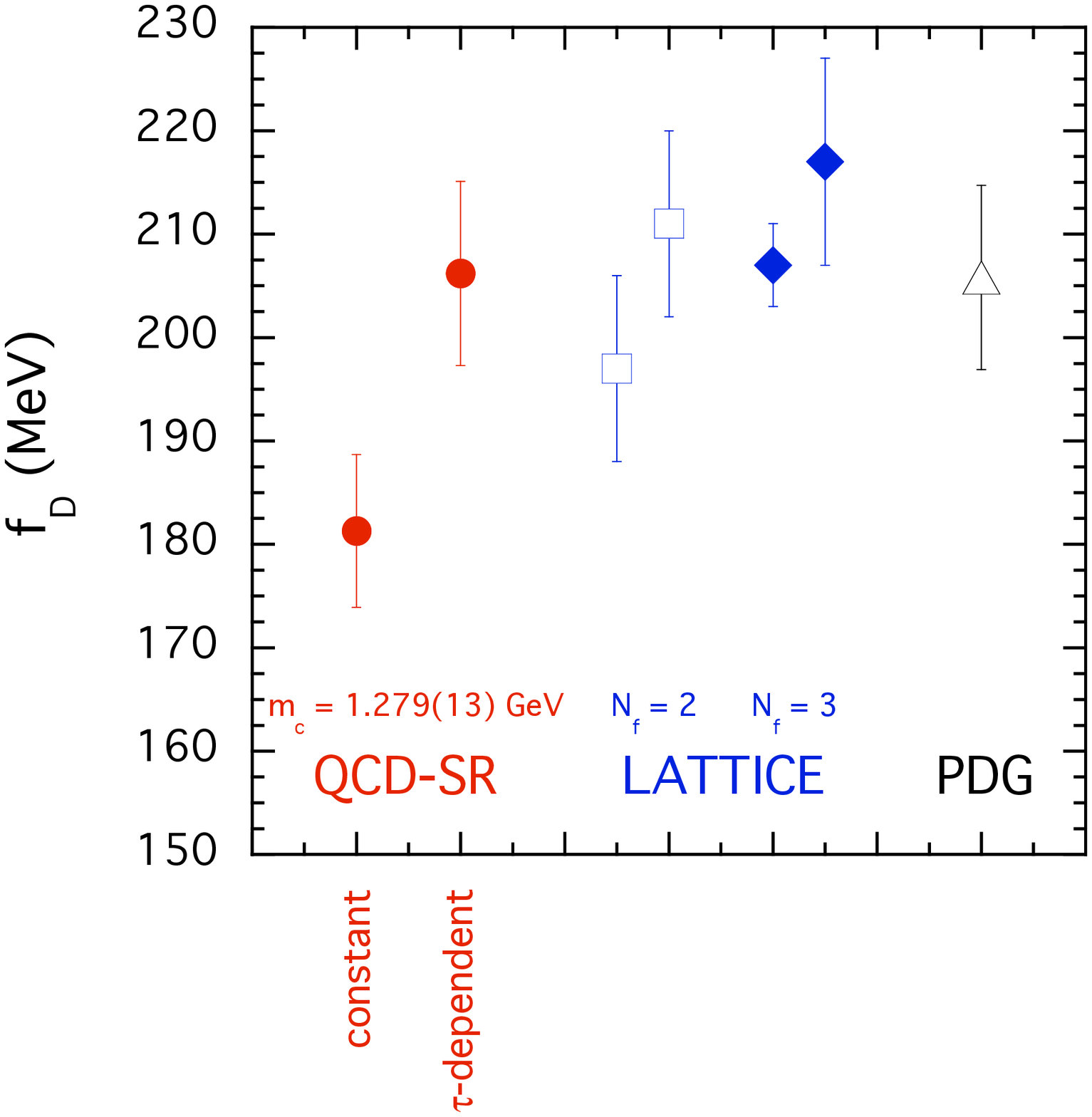}&
\includegraphics[height=6.5cm]{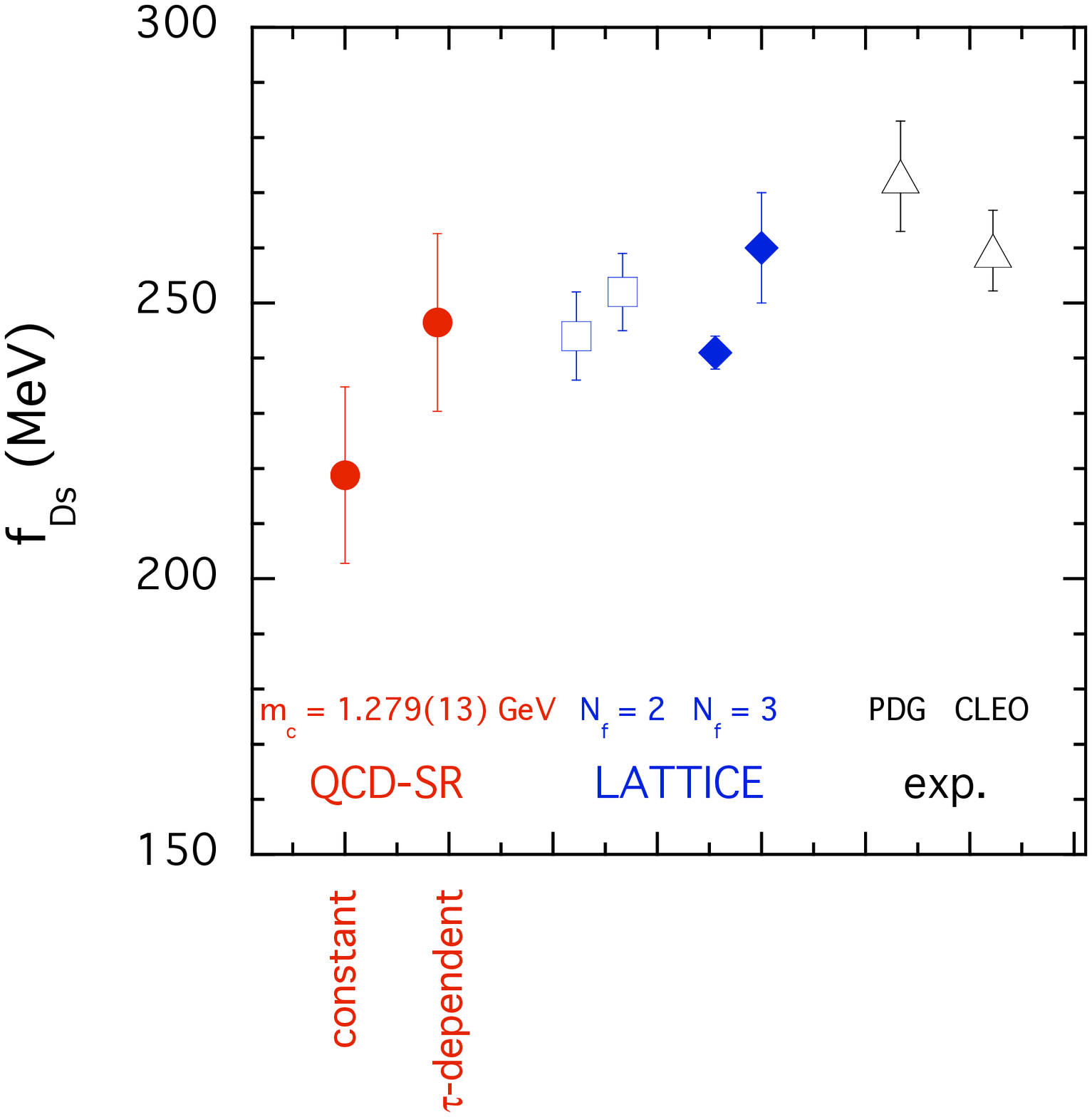}
\\
(c) & (d)
\\
\end{tabular}
\caption{\label{Plot:Dresults}Our results for $f_D$ and $f_{D_s}$:
the bootstrap analysis of the OPE uncertainties (a,b); comparison
with both lattice determinations and experimental data (c,d). For
a detailed list of references, see Ref.~\cite{lms2010}.}
\end{figure}

Applying our extraction procedures to the charmed mesons yields
the following~results:
\begin{eqnarray}
\label{Dresults} f_{D} &=& (206.2 \pm 7.3_{\rm (OPE)} \pm 5.1_{\rm
(syst)})\; \mbox{MeV}, \\ f_{D_s} &=& (245.3 \pm 15.7_{\rm (OPE)}
\pm 4.5_{\rm (syst)})\; {\rm MeV}.
\end{eqnarray}
The OPE error is obtained by the bootstrap allowing for variation
of the QCD parameters (quark masses, $\alpha_s$, condensates) in
the relevant ranges. One observes a perfect agreement of our
predictions with the lattice computations
(Fig.~\ref{Plot:Dresults}). It should be emphasized that~the
$\tau$-dependent threshold constitutes a crucial ingredient for
the successful extraction of the decay constant from the sum rule.
Obviously, the standard $\tau$-independent approximation yields a
much lower value for $f_D$ which lies rather far from the data and
the lattice~results.

\section{Decay constants of the $B$ and $B_s$ mesons}
\begin{figure}[!h]
\begin{tabular}{cc}
\includegraphics[height=6.5cm]{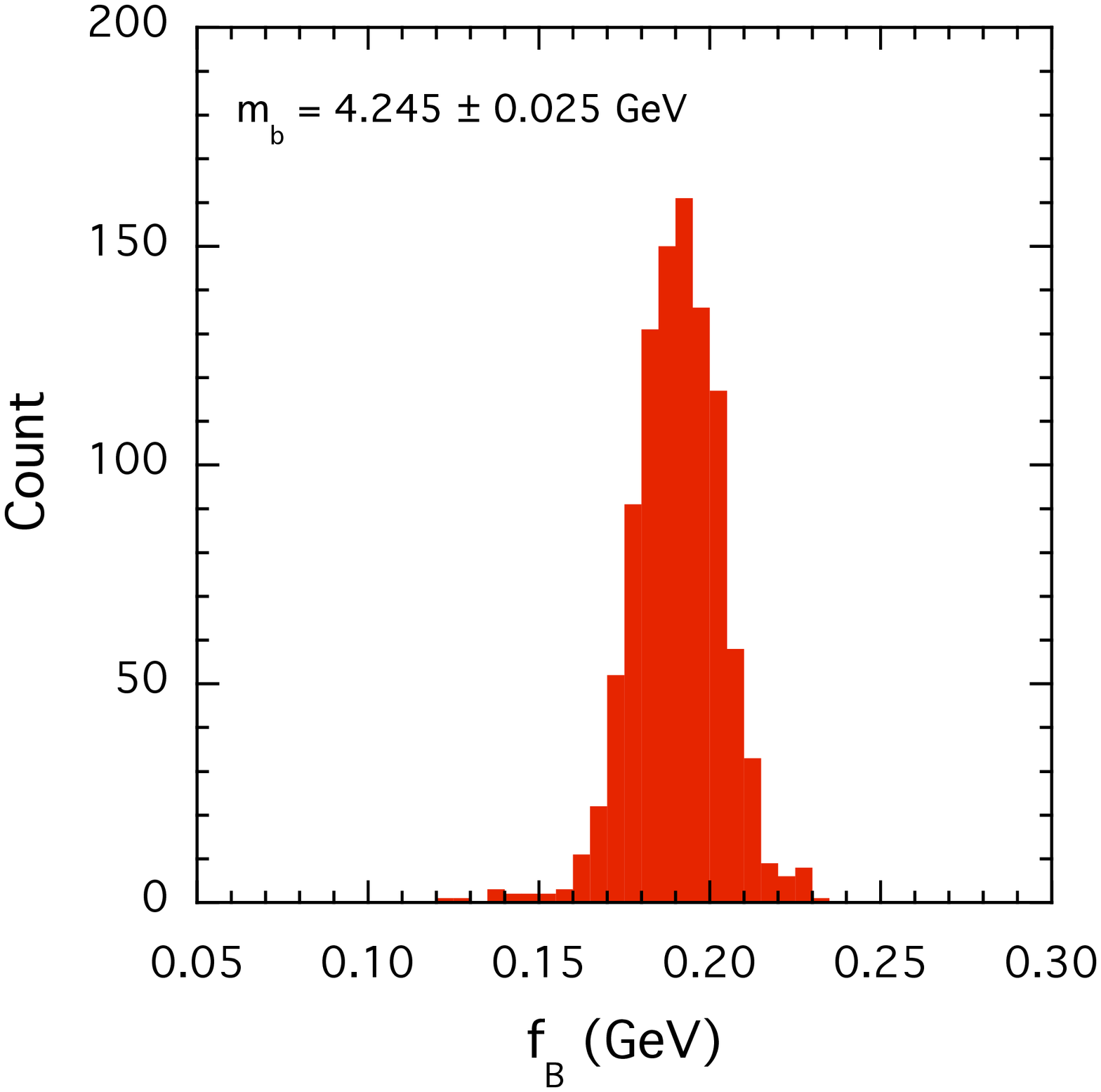}&
\includegraphics[height=6.5cm]{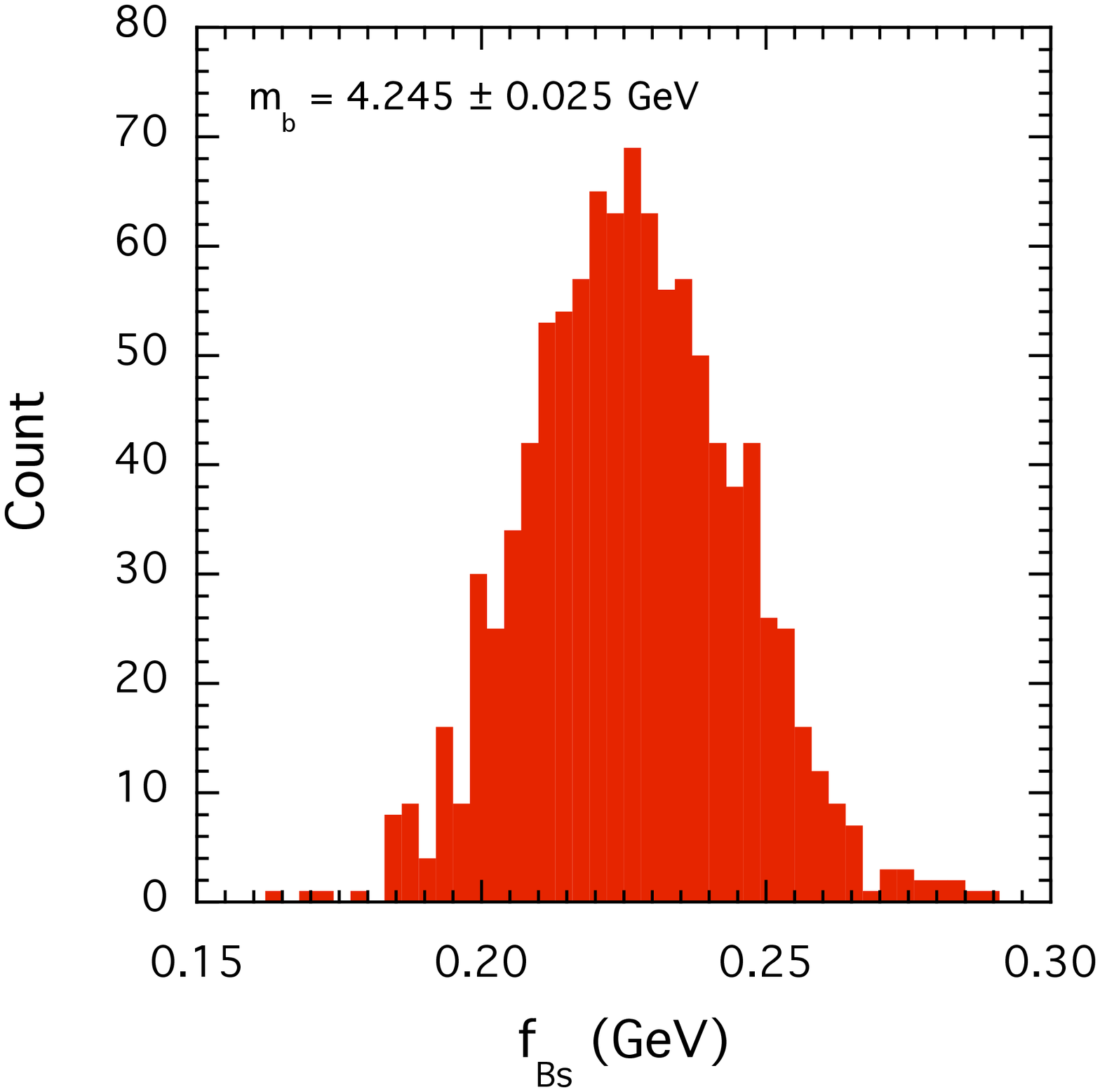}
\\
(a) & (b)
\\
\includegraphics[height=6.5cm]{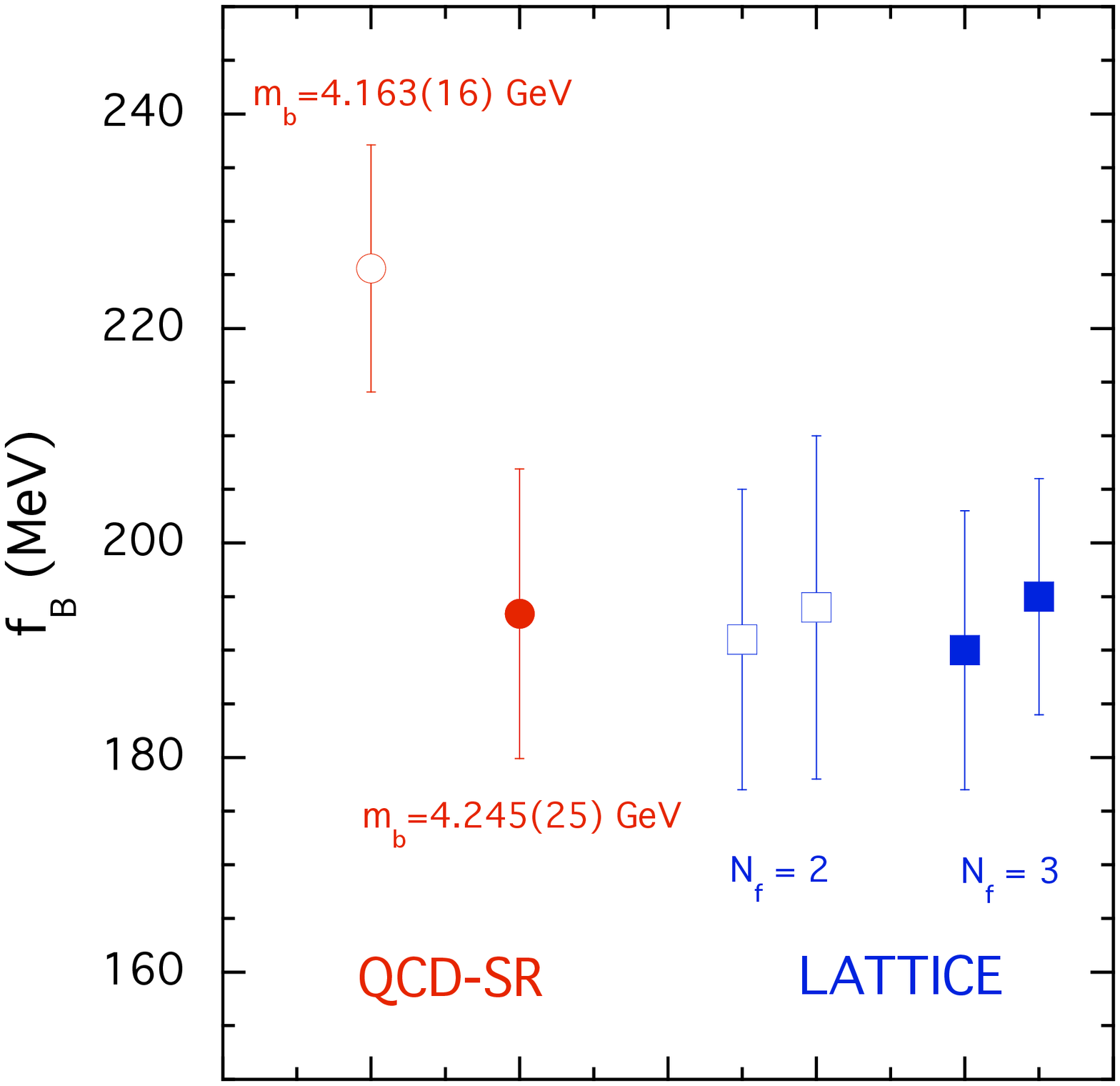}&
\includegraphics[height=6.5cm]{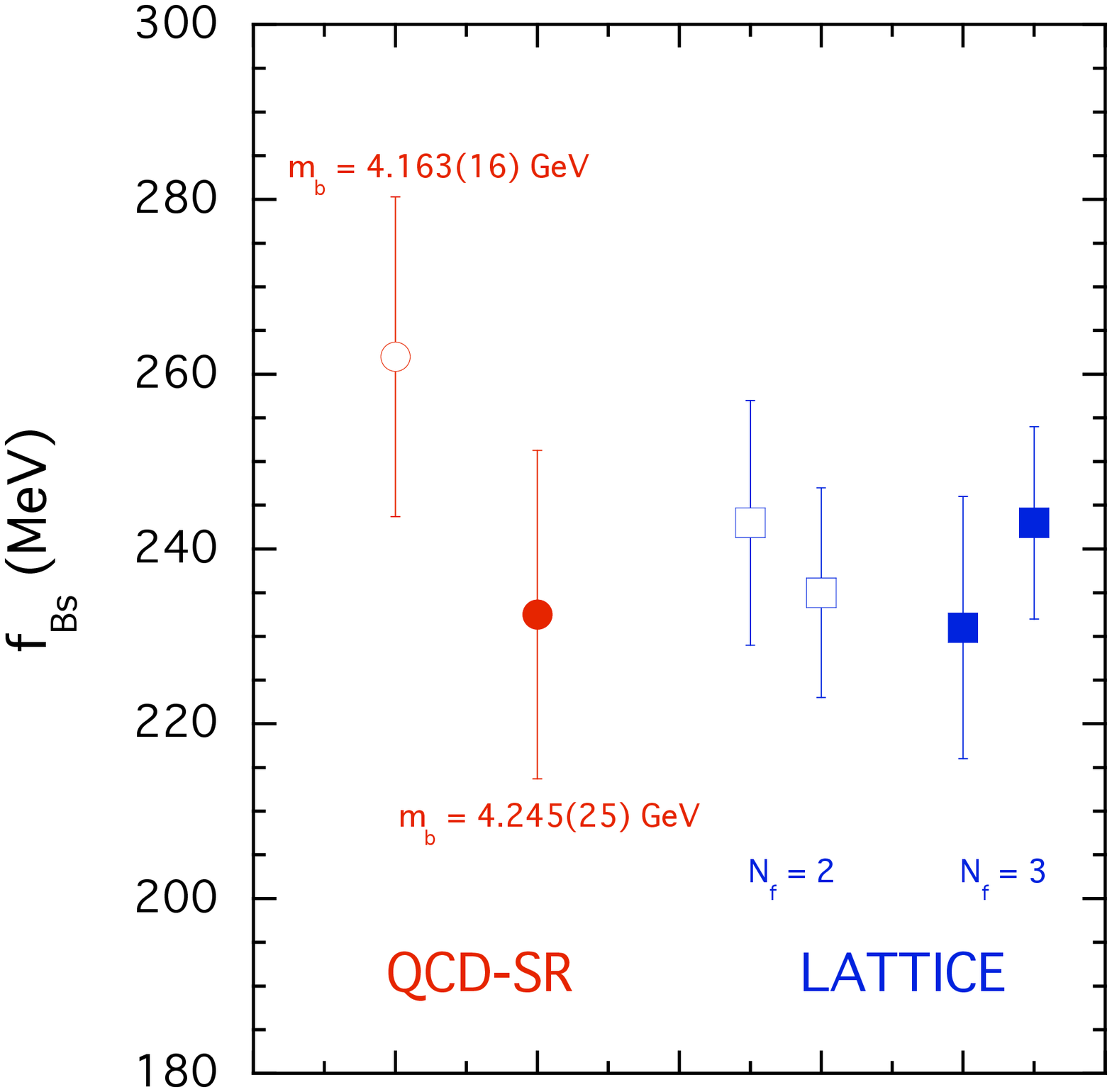}
\\
(c) & (d)
\end{tabular}
\caption{\label{Plot:Bresults}Our results for $f_B$ and $f_{B_s}$:
the bootstrap analysis of the OPE uncertainties (a,b); comparison
with both lattice determinations and experimental data (c,d). For
a detailed list of references, see Ref.~\cite{lms2010}.}
\end{figure}

The values of the beauty-meson decay constants extracted from the
sum rule (\ref{SR_QCD}) turn out to be extremely sensitive to the
precise value of $m_b\equiv \overline{m}_b(\overline{m}_b)$. Our
results for these~decay constants may be parameterized as follows:
\begin{eqnarray}
f_{B}(m_b)&=& \left[193.4 - 37\left(\frac{m_b-\mbox{4.245
GeV}}{\mbox{0.1 GeV}}\right)\right] \mbox{MeV},\\ f_{B_s}(m_b)&=&
\left[232.5 -43\left(\frac{m_b-\mbox{4.245 GeV}}{\mbox{0.1
GeV}}\right)\right] \mbox{MeV}.
\end{eqnarray}
Making use of the $b$-mass range
$\overline{m}_b(\overline{m}_b)=(4.163\pm 0.016)$ GeV \cite{mb}
yields results barely compatible with the lattice determination
(Fig.~\ref{Plot:Bresults}). Requiring the sum-rule
result~for~$f_B$~to

\noindent match the average of the lattice results yields a rather
precise value of the $b$-quark mass:
\begin{eqnarray}
\overline{m}_b(\overline{m}_b)=(4.245\pm 0.025)\; {\rm GeV}.
\end{eqnarray}
For this value of $m_b$, our sum-rule estimates for the $B$- and
$B_s$-meson decay constants~are
\begin{eqnarray}
\label{Bresults} f_{B} &=&(193.4 \pm 12.3_{\rm (OPE)}\pm 4.3_{\rm
(syst)})\; {\rm MeV},\\ f_{B_s} &=& (232.5 \pm 18.6_{\rm (OPE)}\pm
2.4_{\rm (syst)})\; {\rm MeV}.
\end{eqnarray}

\section{Conclusions}
We presented the results of our recent analysis of the heavy-meson
decay constants from the correlator of pseudoscalar currents
\cite{lms2010}. Our special emphasis was laid on the study~of {\em
all\/} uncertainties in the extracted value of the decay constant,
viz., on the OPE uncertainty related to the not precisely known
QCD parameters and on the intrinsic uncertainty of~the method
related to a limited accuracy of the extraction procedure.
According to our recent findings, the accuracy of the sum-rule
predictions may be considerably improved and the intrinsic
uncertainties of hadron parameters may be probed by allowing for
$\tau$-dependent Ans\"atze for the effective continuum threshold.
The parameters of this effective threshold can be fixed by
minimizing the deviation of the dual mass from the known
meson~mass~in the window. This strategy has now been applied to
the decay constants of heavy mesons.

\vspace{.12cm} Our main results are as follows:

\vspace{.12cm} 1. The analysis of charm mesons unambiguously
demonstrates that the application of the Borel-parameter-dependent
effective threshold leads to two essential improvements: (i) the
accuracy of decay constants extracted from the sum rule is
considerably improved; (ii) it has become possible to derive
realistic systematic uncertainties and to reduce their values to
the level of a few percent. By application of our novel extraction
procedures~the results from QCD sum rules have been brought into a
perfect agreement with the findings from lattice QCD and
experiment.

\vspace{.12cm} 2. The $B$ and $B_s$ decay constants are extremely
sensitive to the precise value of
$\overline{m}_b(\overline{m}_b)$. Therefore, no reasonable
predictions for the decay constants may be extracted unless the
$b$-quark mass is known with very high accuracy. On the other
hand, the strong sensitivity of these decay constant to the
$b$-quark mass opens the promising possibility to extract~the
$b$-quark mass if the decay constant is known. Following this line
and matching the results from QCD sum rules for $f_B$ to the
average of the lattice evaluations allows us to obtain a rather
accurate estimate of the $b$-quark mass. Our value is in good
agreement with several lattice results but, interestingly, it does
not overlap with the recent accurate determination of
$\overline{m}_b(\overline{m}_b)$ reported in \cite{mb} (see also
[1]). Clearly, this issue requires further investigation.

\begin{theacknowledgments}
We would like to thank the Organizing Committee for creating and
maintaining for many years such particularly nice and friendly
working atmosphere at this series of workshops. We are grateful to
M.~Jamin for useful discussions at initial stages of this work.
D.M.~was supported by the Austrian Science Fund (FWF) under
project no.~P20573.
\end{theacknowledgments}

\bibliographystyle{aipproc}

\begin{thebibliography}{9}
\bibitem{lms2010}
W.~Lucha, D.~Melikhov, and S.~Simula, arXiv:1008.2698 [hep-ph].
\bibitem{svz}
M.~Shifman, A.~Vainshtein, and V.~Zakharov, \emph{Nucl.~Phys.~B\/}
\textbf{147}, 385 (1979).
\bibitem{chetyrkin}
K.~G.~Chetyrkin and M.~Steinhauser, \emph{Phys.~Lett.~B\/}
\textbf{502}, 104 (2001); \emph{Eur.~Phys.~J.~C\/} \textbf{21},
319 (2001).
\bibitem{jamin}
M.~Jamin and B.~O.~Lange, \emph{Phys.~Rev.~D\/} \textbf{65},
056005 (2002).
\bibitem{lms_2ptsr}
W.~Lucha, D.~Melikhov, and S.~Simula, \emph{Phys.~Rev.~D\/}
\textbf{76}, 036002 (2007); \emph{Phys.~Lett.~B\/} \textbf{657},
148 (2007); \emph{Phys.~Atom.~Nucl.\/} \textbf{71}, 1461 (2008).
\bibitem{lms_new}
W.~Lucha, D.~Melikhov, and S.~Simula, \emph{Phys.~Rev.~D\/}
\textbf{79}, 096011 (2009); \emph{J.~Phys.~G\/} \textbf{37},
035003~(2010); W.~Lucha, D.~Melikhov, H.~Sazdjian, and S.~Simula,
\emph{Phys.~Rev.~D\/} \textbf{80}, 114028 (2009).
\bibitem{lms_qcdvsqm}
W.~Lucha, D.~Melikhov, and S.~Simula, \emph{Phys.~Lett.~B\/}
\textbf{687}, 48 (2010); arXiv:1003.1463 [hep-ph], \emph{Phys.\
Atom.~Nucl.\/} \textbf{73} (in print).
\bibitem{mb}
K.~G.~Chetyrkin \emph{et al.\/}, \emph{Phys.~Rev.~D\/}
\textbf{80}, 074010 (2009).
\end{thebibliography}
\end{document}